\documentclass[journal]{IEEEtran}

\ifCLASSINFOpdf
\else
\fi
\usepackage{amsmath}    
\usepackage{amssymb}    
\usepackage{bm}         
\usepackage{booktabs}
\usepackage{booktabs}

\usepackage{graphicx}
\usepackage{cite}
\usepackage[colorlinks=true, allcolors=black]{hyperref}

\usepackage[capitalise,noabbrev,nameinlink]{cleveref}
\crefname{figure}{Fig.}{Figs.}
\Crefname{figure}{Figure}{Figures}
\crefname{equation}{Eq.}{Eqs.}
\Crefname{equation}{Equation}{Equations}
\crefname{table}{Table}{Tables}
\Crefname{table}{Table}{Tables}
\crefname{section}{Sec.}{Secs.}
\Crefname{section}{Section}{Sections}

\IEEEaftertitletext{\vspace{-1.5\baselineskip}} 

\usepackage{graphicx}
\usepackage{placeins}
\usepackage{url}

\begin{document}
\title{Generalized Nyquist Criterion Limitations and Misconceptions for Frequency Domain Stability Analysis of Inverter-based Resources Integrated Power Grids}
\author{Hassan~Yazdani,~\IEEEmembership{Graduate Student Member,~IEEE,}~Saeed~Lotfifard,~\IEEEmembership{Senior~Member,~IEEE}

\thanks{H. Yazdani and S. Lotfifard are with the School of Electrical Engineering and Computer Science, Washington State University, Pullman, WA 99164 USA (e-mail: hassan.yazdani@wsu.edu; s.lotfifard@wsu.edu).}%
}
\maketitle
\begin{abstract}
This paper presents theoretical and numerical studies demonstrating the limitations of the Generalized Nyquist Criterion (GNC) in assessing the small-signal stability of inverter-based resources (IBRs) modeled as multi-input, multi-output (MIMO) systems. GNC leads to unnecessary computational burden and increased analytical complexity for nominal stability analysis. The paper demonstrates that the GNC framework is not reliable for robust-stability analysis and may lead to misleading results when applied to MIMO IBR systems. To demonstrate these limitations and enable a proper MIMO robust stability assessment of IBRs using µ-analysis, the paper develops a model-uncertainty-augmented representation of an IBR-integrated power grid. The proposed model explicitly captures the structural characteristics and spatial distribution of various uncertainty sources, such as parametric variations, unmodeled dynamics, and measurement errors. Since these uncertainties naturally occur in the physical three-phase (abc) systems, they are systematically transformed into the dq0 frame within the developed model. 
\end{abstract}


\section{Introduction}
\IEEEPARstart{S}{tability} analysis of IBRs is crucial for their grid integration studies. Therefore, various methods have been proposed as reviewed in [1-8]. A common approach is the impedance-based method using GNC. The premise of this approach is to extend the frequency-domain stability analysis methods commonly used in single-input single-output (SISO) systems to MIMO systems. While frequency-domain methods are effective tools for nominal and robust stability analysis in SISO systems, unfortunately, most advantages cannot be extended to MIMO systems. The focus and contributions of this paper are as follows:

\begin{itemize}
    \item\emph{Demonstrating the limitations of GNC for MIMO IBRs’ stability analysis:} This paper discusses and numerically demonstrates the inherent limitations of the GNC framework for both nominal and robust stability of IBRs. It highlights that GNC is not appropriate for nominal stability testing because it leads to unnecessary computational burden and increased analytical complexity. It demonstrates that GNC should be avoided for robust stability assessment of IBRs, as it may provide unreliable results. 
    \item\emph{Developing a model-uncertainty-augmented representation for proper MIMO IBRs stability analysis:} To enable proper robust stability analysis of MIMO IBR systems, a critical requirement is the use of models that accurately represent the system uncertainties.
\end{itemize}

In [9-10], the utilized models for robustness analysis consider only parametric uncertainties. In [11-12], uncertainties are modeled as lumped, diagonal, output multiplicative perturbations, which overlook the specific locations where uncertainties arise and ignore the possible off-diagonal (coupled) interactions. In [13], more detailed uncertain models are developed. However, the uncertainties are directly applied to the frame, while real-world uncertainties arise in physical three-phase systems. To overcome these limitations and enable proper robust stability assessments, this paper develops a model-uncertainty-augmented representation of an IBR-integrated system. The developed model captures various sources of uncertainty and explicitly represents their structural characteristics and spatial distribution within the system. The uncertainties in the three-phase systems are systematically converted into the frame of the developed model.   

The remainder of this paper is organized as follows. Section. II presents the theoretical analysis of the limitations, and Section. III explains the proposed uncertainty-augmented model and presents the numerical analysis of the limitations. Finally, conclusions are presented in Section. IV.

\section{Theoretical Analysis of Limitations of Generalized Nyquist Criterion for IBRs}
\subsection{Some background information on nominal stability analysis and robust stability assessment }
Fig. 1 shows a typical closed-loop dynamical SISO system. Assume $g(s)=\frac{n(s)}{d(s)}$ is a plant transfer function. In Fig. 1, assuming an uncertainty block $u(s)=1$, the nominal closed-loop transfer function is $\frac{g(s)}{1+g(s)}=1-\frac{1}{1+g(s)}$ . If the goal is to test the nominal stability of the system, one can directly check if  $\frac{1}{1+g(s)}$ does not have any closed right half-plane poles, or equivalently $n(s)+d(s)$ has all zeros on the open left half-plane.  This nominal stability test can be efficiently conducted using the Routh-Hurwitz stability criterion [14]. This approach does not require Nyquist or Bode plots. In certain scenarios, the frequency-domain model, such as the $s$-domain transfer function, of a device may be unavailable. For example, in black-box models, only frequency-response data (e.g., Bode plots) may be available. In a single-IBR connected to an infinite bus system, frequency-domain stability analysis methods can be directly applied to the collected frequency-response data. 

\begin{figure}[!t]
  \centering
  \includegraphics[width=\linewidth]{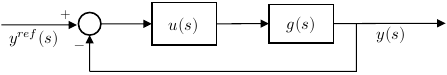}
  \caption{A typical structure of a SISO closed-loop system with an uncertainty block $u(s)$.}
\end{figure}

For SISO systems, tools such as the Bode and Nyquist criteria are effective because they provide insight into both nominal and robust stability by evaluating gain margin (GM) and phase margin (PM) without requiring explicit frequency-response models, such as $s$-domain transfer functions. However, these advantages do not extend to MIMO systems. As discussed later, GM and PM in the GNC are not reliable indicators for robust stability analysis in MIMO systems. Although the GNC yields correct results for nominal stability analysis of MIMO systems, it was not originally intended for such applications, and simpler, more effective approaches are available. Specifically, the transfer function of an individual inverter-based resource (IBR) can be reconstructed from measured frequency response data using vector-fitting techniques [15-16]. For multi-IBR systems, direct use of frequency-response data becomes increasingly cumbersome and does not provide valuable information about interactions among IBRs, such as participation factors [15]. Reference [17] presents a stability analysis of the multi-IBR system based on frequency-response models, specifically, $s$-domain transfer matrices. In contrast, [15][18] convert these $s$-domain transfer matrices of individual IBRs into state-space representations and perform the stability analysis of the overall system in the state-space form. State-space modeling offers significant advantages for controller design. For nominal stability assessment, state-space modeling and $s$-domain transfer-matrix modeling yield equivalent results in systems with minimum realization. However, the state-space formulation provides several advantages over $s$-domain transfer functions by extending beyond their restrictive input-output representation for stability analysis. It offers additional structural insight by linking system dynamics to state variables associated with specific physical points or components within the system, which is critical for root-cause analysis of dynamic behavior. Furthermore, unlike transfer function-based approaches, which may obscure internal modes through pole-zero cancellations in systems with non-minimum realizations, the state-space representation preserves the complete system dynamics.     

The GNC framework was originally developed for robustness assessment within the loop-shaping controller design process, aiming to extend SISO loop-shaping controller design to MIMO systems. As will be discussed later, it was subsequently recognized that robustness analysis in the GNC framework does not always provide reliable results. Consequently, its use has been avoided over the past several decades and was not adopted by the mainstream control systems community. Instead, methodologies specifically designed for MIMO systems, such as $\mu$-analysis, have been developed.

In SISO systems, GM and PM quantify permissible variation in gain or phase that the open-loop system can withstand at its crossover frequencies before instability occurs. These indices are effective tools for loop-shaping controller design. They can be interpreted as measures of the allowable multiplicative uncertainty in the open-loop transfer function near that frequency. However, robustness characterization based on GM and PM is valid only when the uncertainty does not substantially alter the overall frequency response; otherwise, modern robustness analysis methods are recommended [9].

As shown in Fig. 1, GM and PM represent the perturbation in the form of $u(s)g(s)$ or $g(s)u(s)$ where $u(s)=ke^{j\theta}$. For $\theta=0$, the parameter $k$ acts as a scaling factor that determines how much the Nyquist plot of $g(j\omega)$ can be scaled before the system becomes unstable (i.e., the GM). For $k=1$, $\theta$ represents how much the Nyquist plot of $g(j\omega)$  can be rotated before the system becomes unstable (i.e., the PM). It is important to note that the definitions of GM and PM assume that one quantity varies while the other remains constant, an assumption that may not hold in real-world systems. In most practical scenarios, uncertainty in SISO systems influences both phase and gain. A system with appropriately designed phase and gain margins can typically accommodate these simultaneous variations while maintaining satisfactory stability. Consequently, phase and gain margins offer valuable engineering insight and continue to serve as effective tools for controller design in SISO systems using loop-shaping methods. It should be noted that in very rare situations, typically in mathematically constructed cases known as pathological examples, as discussed in [19], [20, pages 10-17] [21, page 375], a system may fail to exhibit adequate robustness despite possessing acceptable GM and PM. In some controller design and analysis methods for IBRs [22-23], disk margin is used, which considers simultaneous phase and gain perturbation. Also, other names used instead of disk margins are vector margin [24, page 124] and modulus margin [25, page 70]. Delay margin, which is related to phase margin, is also defined in [25, page 70]. Nevertheless, in general, phase and gain margins remain highly effective and intuitive tools for robustness assessment and controller design of SISO systems. Moreover, in SISO systems, gain and phase margins are directly linked to controller performance and are commonly used as design criteria [14]. However, as will be discussed later, these concepts do not generalize to MIMO systems.

\subsection{Generalized Nyquist Criterion}

Early efforts in robust stability analysis of MIMO systems sought to extend the methods developed for SISO systems to MIMO systems. One such method was the “single-loop-at-a-time” approach, which has also been applied in the context of IBRs’ stability analysis [26]. However, [27] presented counterexamples illustrating the limitations of using single-loop analysis in MIMO systems. Likewise, [28] and [29] have also highlighted that calculating GM and PM on a per-loop basis is not appropriate for MIMO systems. In [30-31], the conditions under which the SISO approximation of IBRs, neglecting the influence of off-diagonal elements, leads to inaccurate results are discussed.

\begin{figure}[!t]
  \centering
  \includegraphics[width=\linewidth]{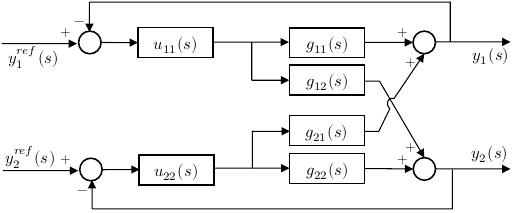}
  \caption{A typical structure of a two-input two-output (TITO) system $\mathbf{G}(s)$ with a diagonal input uncertainty block $\mathbf{U}(s)$.}
\end{figure}

The GNC aims to extend the classical Nyquist stability test to MIMO systems, providing a systematic framework in multi-loop feedback configurations. In particular, the GM and PM concepts in SISO systems enable loop-shaping control design, which is a very effective approach for SISO systems. GNC intended to devise a systematic approach to enable loop-shaping controller design for MIMO systems, such as [32]. However, GNC can extend SISO GM and PM concepts to MIMO systems only for a special case of uncertainty called the uniform uncertainty condition, where the disturbance is identical across all channels, and there is no cross‑channel coupling. For example, Fig. 2 shows a MIMO system with two inputs and two outputs (TITO). In Fig. 2, a diagonally structured uncertainty matrix $\mathbf{U}(s)$ with identical diagonal elements is considered as follows in (1).
\begin{equation}
\label{eq1}
\mathbf{U}(s)
=
\begin{bmatrix}
u_{11}(s) & 0\\
0 & u_{22}(s)
\end{bmatrix},
\qquad
\text{with }
u_{11}(s)=u_{22}(s)=k e^{j\theta}.
\end{equation}
In other words, $\mathbf{U}(s)$ is a scalar matrix as in (2).
\begin{equation}
\label{eq:transformation_matrix_identity}
\mathbf{U}(s)
=
\begin{bmatrix}
u_{11}(s) & 0\\
0 & u_{22}(s)
\end{bmatrix}
=
k e^{j\theta}
\begin{bmatrix}
1 & 0\\
0 & 1
\end{bmatrix}
=
k e^{j\theta}\mathbf{I}_{2\times 2}.
\end{equation}

In robust stability analysis, the nominal system is assumed stable, and the goal is to measure how far it is from losing stability. A direct approach would require checking, for every admissible perturbation $\mathbf{U}(s)$, whether the determinant condition $\text{det}(\mathbf{I}+\mathbf{G}(s)\mathbf{U}(s))=0$ admits right half-plane zeros corresponding to unstable poles of the closed-loop system, and $\mathbf{I}$ is the identity matrix. Since this involves checking infinitely many perturbations, the task is prohibitively complex. The GNC offers a more tractable alternative by examining the loci of the characteristic function on the imaginary axis $s=j\omega$. Because the determinant is a nonlinear operator, the expression $\text{det}(\mathbf{I}+\mathbf{G}(j\omega)\mathbf{U}(j\omega))$ depends nonlinearly on the perturbation parameter $\mathbf{U}(j\omega)$ as follows [33, page 60]. 
\begin{equation}
\label{eq:characteristic_determinant}
\begin{aligned}
&\det\!\left(
\mathbf{I}_{2\times2}
+\mathbf{G}(j\omega)\mathbf{U}(j\omega)
\right)
\\
&\quad =
\det\!\left(
\mathbf{I}_{2\times2}
+k e^{j\theta}
\begin{bmatrix}
g_{11}(j\omega) & g_{12}(j\omega)\\
g_{21}(j\omega) & g_{22}(j\omega)
\end{bmatrix}
\right).
\end{aligned}
\end{equation}
As a result, the gain and phase margins cannot be obtained from the Nyquist plot of $\text{det}(\mathbf{I}+\mathbf{G}(j\omega)\mathbf{U}(j\omega))$. To address this difficulty, the GNC does not rely on encirclements of the origin by the Nyquist plot of $\text{det}(\mathbf{I}+\mathbf{G}(j\omega)\mathbf{U}(j\omega))$. Instead, the equality (4) is used. 
\begin{equation}
\label{eq:characteristic_determinant_eigenvalues}
\begin{aligned}
\det\!\left(
\mathbf{I}+\mathbf{G}(j\omega)\mathbf{U}(j\omega)
\right)
&=
\prod_{i=1}^{2}
\lambda_i\!\left(
\mathbf{I}
+\mathbf{G}(j\omega)k e^{j\theta}\mathbf{I}
\right)
\\
&=
\prod_{i=1}^{2}
\left(
1+k e^{j\theta}\lambda_i\!\left(\mathbf{G}(j\omega)\right)
\right)
\end{aligned}
\end{equation}
where $\lambda_i (\mathbf{G}(j\omega))$ are the eigenvalues of the transfer function matrix $\mathbf{G}(j\omega)$. The problem is reformulated as counting the encirclements of the critical point $-1+j0$ by the locus of $ke^{j\theta} \lambda_i (\mathbf{G}(j\omega))$ which can be visualized by plotting the eigenvalue trajectories $\lambda_i (\mathbf{G}(j\omega))$, scaling them by the gain $k$, and rotating them by the phase angle $\theta$. In other words, the magnitudes of the eigenvalues $\lambda_i (\mathbf{G}(j\omega))$ vary linearly with the gain parameter $k$, and the phases of the eigenvalues $\lambda_i (\mathbf{G}(j\omega))$ vary linearly with the phase parameter $\theta$. Instability may occur if, under perturbations, the number of encirclements of $-1+j0$ changes. Note that the nominal system is assumed to be stable in the absence of uncertainty; instability arises if some eigenvalue trajectory crosses through $-1+j0$, since beyond this crossing the Nyquist encirclement count may change, marking the onset of possible instability. This condition corresponds to a singularity, expressed for SISO systems as $1+g(j\omega)u(j\omega)=0$  or $\text{det}(\mathbf{I}+\mathbf{G}(j\omega)\mathbf{U}(j\omega))=0$ for MIMO systems. Note that (4) does not extend to determinants, i.e., $\text{det}(\mathbf{I}+\mathbf{G}(j\omega)\mathbf{U}(j\omega))\neq1+\text{det}(\mathbf{G}(j\omega)\mathbf{U}(j\omega))$. Thus, when using the Nyquist plot of $\text{det}(\mathbf{I}+\mathbf{G}(j\omega)\mathbf{U}(j\omega))$, encirclements must be checked around the origin and not around $-1+j0$.

Robust stability analysis of MIMO systems using GNC has the following limitations:  
\begin{itemize}
    \item A general and realistic perturbation matrix, for example in a $2\times2$ MIMO system, takes the form $\mathbf{U}(s)=\begin{bmatrix}u_{11} & u_{12}\\ u_{21} & u_{22}\end{bmatrix}$ where the uncertainty in each channel may differ (i.e., $u_{11}\neq u_{22}$) and the channels can also be coupled, i.e., the off diagonal terms $u_{12}$ and $u_{21}$ are generally nonzero. Uniform uncertainty, where $\mathbf{U}(s)$ reduces to a scalar perturbation matrix (2), affects the nominal Nyquist plot in a predictable manner: the nominal Nyquist plot only expands or contracts under uniform gain perturbations and only rotates under uniform phase perturbations. Consequently, the predicted GM and PM by the GNC remain consistent with their actual values in this restricted condition. However, uniform uncertainty is a rare and highly restrictive condition. In practical systems, uncertainty is generally non‑uniform, which may alter the nominal Nyquist plot in unpredictable, nonlinear ways. A non‑uniform gain perturbation matrix does not merely scale the nominal Nyquist plot but may alter its gain/phase values differently across frequencies; likewise, a non‑uniform phase perturbation does not simply rotate the nominal Nyquist plot but may modify the gain and phase differently across frequencies. As a result, GNC‑based gain and phase margins may provide unreliable indications of robustness. In contrast to SISO systems, where predicted and actual gain and phase margins are consistent, this consistency may not hold for MIMO systems under realistic non-uniform uncertainty conditions. Also, unlike in SISO systems, the distance of a given point on the nominal Nyquist plot from the critical point $-1+j0$ is not a reliable indicator of vulnerability to instability in MIMO systems. This is because the shape of the perturbed Nyquist plot can change unpredictably under non uniform uncertainty. 
    \item GNC relies on the direct calculation of eigenvalues $\lambda_i (\mathbf{G}(j\omega))$ to evaluate their proximity to instability/singularity. However, it is well known that eigenvalues are not a proper metric for measuring the distance to singularity. Instead, singular values provide a more appropriate measure of closeness to singularity as illustrated in [34, page 177]. Additionally, [35] provides explanations of why singular value analysis is preferred over eigenvalue analysis for robustness evaluation in MIMO systems. 
\end{itemize}

When implementing the GNC, several key issues must be considered. For instance, proper labeling and indexing of eigenvalues in the GNC framework are crucial. An example of incorrect labeling is presented later in the case study section. This issue is analogous to the eigenvalue tracking and ordering problem in state space eigenvalue analysis [36]. Moreover, as discussed in [37][38], in MIMO systems poles located on the imaginary axis must be treated as if they were RHP poles when applying the GNC. In [39], the implications of this consideration in power system applications are demonstrated. Also, Nyquist plots may exhibit asymmetry about the real axis for irrational functions. To capture the full behavior of the eigenvalue loci, it is essential to plot them over the entire frequency range from $-\infty$ to $+\infty$, rather than from $0$ to $+\infty$. In SISO systems, GM and PM are directly linked to controller performance and are commonly used as design criteria [14]. In contrast, there is no systematic approach for determining the required phase margin in MIMO systems that relates to its performance. 

Nyquist stability analysis is based on the Principle of Argument [14, page 335]. The GNC can be implemented based on $\text{det}(\mathbf{I}+\mathbf{G}(j\omega))$, the eigenvalues $\lambda (\mathbf{I}+\mathbf{G}(j\omega))$, or $\lambda (\mathbf{G}(j\omega))$. In the approach based on $\lambda (\mathbf{G}(j\omega))$, stability is determined by counting the number of encirclements of the point $-1+j0$ made by the eigenvalue loci $\lambda (\mathbf{G}(j\omega))$. This requires examining both the magnitude and the phase of $\lambda (\mathbf{G}(j\omega))$ to determine the number of crossings along the negative real axis on the segment $(-\infty,-1)$. In the approach based on $\lambda (\mathbf{I}+\mathbf{G}(j\omega))$, stability is assessed by counting the number of encirclements of the origin made by $\lambda (\mathbf{I}+\mathbf{G}(j\omega))$. This can be accomplished by analyzing the phase of $\lambda (\mathbf{I}+\mathbf{G}(j\omega))$ and observing the crossings of the negative real axis $(-\infty,0)$.  This strategy is used in [40] for IBRs stability analysis. As discussed before, the reason for using $\lambda (\mathbf{G}(j\omega))$ in the GNC framework is to define GM and PM. 

\begin{figure}[!t]
  \centering
  \includegraphics[width=\linewidth]{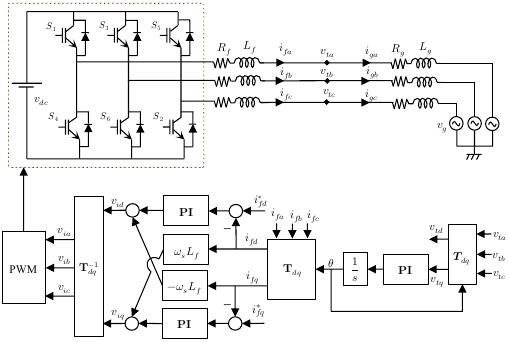}
  \caption{Schematics of a typical grid-connected VSC.}
\end{figure}

\subsection{Different GM and PM concepts for MIMO systems}
Several methods, including those described in [41-42], have applied the concept of forbidden regions to the stability analysis of MIMO systems. In SISO systems, forbidden regions are defined using the Nyquist phase and gain margins, which serve as reliable indicators of robustness [43]. However, extending this concept to MIMO systems presents significant challenges. In contrast to the SISO case, Nyquist-based phase and gain margins in MIMO systems do not provide reliable measures of robustness and should not serve as the basis for defining forbidden regions.

To address the limitations of GM and PM in the GNC framework, different methods have been proposed. In [44], GM and PM based on singular values are defined. The method presented in [44] is further extended in [45]. In [46-47], singular value-based GM and PM are also developed. Principal gains and principal phases concepts are discussed in [48]. In [49], the robustness test metrics of [44] are applied to the stability analysis of wind turbines. As discussed in [34, page 179], the GM and PM calculated using [44] are very conservative, since the method does not assume any particular structure in the perturbation matrix $\mathbf{U}(s)$, and instead evaluates the worst-case scenario that may not happen in real-world systems. However, the results are “reliable”, as they consider all possible perturbation configurations. In contrast, margins calculated using the GNC assume a uniform diagonal structure for the uncertainty matrix. This makes GNC-based margins less conservative, but also unreliable as discussed before. 

In [50-52], non-uniform loop phase and gain margins using optimization methods have been proposed to explicitly enforce the structure of independent loop phase and gain margins in the perturbation matrix $\mathbf{U}(s)$, where $\mathbf{U}(s)$ is assumed to have only diagonal elements of the form $k_i e^{j\theta_i}$, each potentially with different values.  These approaches aim to identify the smallest perturbation that can destabilize the system. In this context, a perturbation magnitude of one and an angle of zero corresponds to the nominal system, so the objective is to find the minimum deviation from one that leads to instability. However, these methods suffer from different limitations, including: (a) According to [32], the robustness assessments obtained using these techniques are not reliable, as the presence of off-diagonal perturbations is not considered. (b) As also highlighted in [53, page 19], the underlying optimization problems are non-convex, which introduces the risk of convergence to a local minima. As a result, the computed GM and PMs may not reflect the true worst-case margins; smaller perturbations capable of destabilizing the system may exist but remain undetected. An application of the concept introduced in [50] to TITO systems is presented in [54]. 

In [53][55-56], methods are proposed that do not rely on traditional loop gain and phase margins. Instead, they introduce matrix-based representations of GM and PM. They argue that, in SISO systems, margins can be represented as $ke^{j\theta}$, where $k$ is a positive scalar (analogous to a positive definite matrix in MIMO systems), and $e^{j\theta}$  captures the phase rotation (analogous to a unitary matrix in MIMO systems). However, the limitations of this matrix-based definition of GM and PM are discussed in [52]. One major drawback is that the results may lack physical interpretability, potentially leading to nonrealistic scenarios that are unlikely to occur in practical systems. 

Various methods, such as [28][57], have been proposed to incorporate information about the uncertainty structure into the perturbation matrix $\mathbf{U}(s)$, aiming to reduce the conservatism inherent in traditional singular value analysis. However, these methods lack the flexibility to fully exploit all available information about the structure and magnitude of uncertainties. In contrast, MIMO robustness tools such as $\mu$-analysis can readily incorporate such information in the analysis. In the $\mu$-analysis, the robustness is quantified using the structured singular value (SSV) denoted by $\mu_\mathbf{\Delta}$, where the system is stable for $\mu_\mathbf{\Delta} \leq 1$ [58]. In the next section, the robustness analysis of IBRs using $\mu$-analysis will be further discussed.
\begin{figure}[!t]
  \centering

  \includegraphics[width=\linewidth]{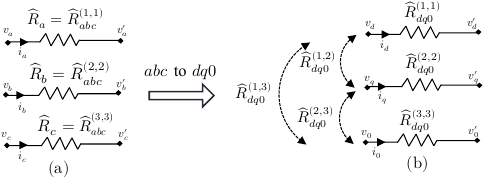}
  \caption{$abc$-to-$dq0$ transformation of uncertain filter resistances.}

  \vspace{0.5em}

  \includegraphics[width=0.6\linewidth]{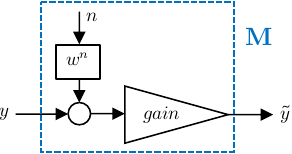}
  \caption{Measurement model representation.}

\end{figure}
\section{Numerical Demonstration of Limitations of GNC for IBRs}

In this section, first the model uncertainty-augmented representation of the system is developed. This model provides insight into how uncertainties may arise and influence the stability of the system. Then, it will be employed for evaluating
the GNC, and a proper MIMO robust stability analysis using $\mu$-analysis is considered. Fig. 3 shows a typical grid-connected voltage source converter (VSC) that represents a grid‑following configuration. In this study, the model is used exclusively to demonstrate the concepts discussed numerically. The analyses can be extended to systems with alternative control structures, including other grid-forming controllers, although such extensions are beyond the present scope. The models described in the following sections represent the linearized system dynamics around the operating point, which is a standard approach in small-signal stability analysis.

\subsection{Model-uncertainty-augmented system}
\textit{Parametric uncertainty:} The uncertain parameters denoted by the set $p \in \{p_1,\ldots,P\}$ can be expressed as follows:
\begin{equation}
\label{eq:uncertain_parameter}
\widehat{p}
=
p_0 \times \left(1+w_p\delta_p\right)
\end{equation}
where $p_0$ is the mean value, given by $p_0=\frac{p_min+p_max}{2}$, $w_p$ represents the relative uncertainty, and $\delta_p$ is a real, scalar, and random parameter bounded by $\lvert\delta_p\rvert\leq1$. This type of uncertainty may appear at various system parameters, such as the inverter filter parameters. 

It is important to highlight one of the most significant advantages of MIMO robust stability analysis methods, such as $\mu$-analysis: their ability to explicitly incorporate information about the structure of system uncertainties. For instance, VSC parameters, such as $L_f$ and $R_f$ in Fig. 3, may experience less uncertainty and variation compared to those of the power grid equivalent model, such as $L_g$ and $R_g$, which may vary due to grid topology changes and outages. This information regarding the relative accuracy of the parameter values can be readily incorporated into the robustness analysis by assigning larger deviations to the grid parameters relative to those of the VSC in the model. However, this cannot be incorporated in general into GNC GM and PM calculations.

Uncertain parameters naturally appear in the original physical frame (i.e., $abc$ frame). In this paper, the following procedure is developed to transform them into the $dq0$ frame, which will be used later for developing the model-uncertainty-augmented representation of the grid-tied VSC system in $dq0$ frame. For the resistance, according to Fig. 4-(a), (6) holds,
\begin{equation}
\label{eq:abc_voltage_relation}
\begin{bmatrix}
v_a-v_a'\\
v_b-v_b'\\
v_c-v_c'
\end{bmatrix}
=
\widehat{\mathbf{R}}_{abc}
\begin{bmatrix}
i_a\\
i_b\\
i_c
\end{bmatrix}
\end{equation}
where
\begin{equation}
\label{eq:abc_uncertain_resistance}
\widehat{\mathbf{R}}_{abc}
=
R_0
\begin{bmatrix}
1+w_{R_a}\delta_{R_a} & 0 & 0\\
0 & 1+w_{R_b}\delta_{R_b} & 0\\
0 & 0 & 1+w_{R_c}\delta_{R_c}
\end{bmatrix}
\end{equation}
By replacing $i_{dq0}=\mathbf{T}_{dq0}i_{abc}$ and $v_{dq0}=\mathbf{T}_{dq0}v_{abc}$ into (6), the following in (8) is derived,
\begin{equation}
\label{eq:dq0_voltage_relation}
\begin{bmatrix}
v_d-v_d'\\
v_q-v_q'\\
v_0-v_0'
\end{bmatrix}
=
\mathbf{T}_{dq0}
\widehat{\mathbf{R}}_{abc}
\mathbf{T}_{dq0}^{-1}
\begin{bmatrix}
i_d\\
i_q\\
i_0
\end{bmatrix}
=
\widehat{\mathbf{R}}_{dq0}
\begin{bmatrix}
i_d\\
i_q\\
i_0
\end{bmatrix}.
\end{equation}
where $\widehat{\mathbf{R}}_{dq0}$ is the resistance matrix in $dq0$ frame and $\mathbf{T}_{dq0}$ is the Park transformation matrix transforming the $abc$ quantities to $dq0$ the system frame. This means that the decoupled $\widehat{\mathbf{R}}_{abc}$, when transformed into the $dq0$ frame, results in a coupled $\widehat{\mathbf{R}}_{dq0}$, as shown in Fig. 5-(b). However, for simplicity, if the resistance values in the $abc$ frame are assumed to be identical, meaning in (7) the uncertainties satisfy $w_{R_a}\delta_{R_a}=w_{R_b}\delta_{R_b}=w_{R_c}\delta_{R_c}$, then $\widehat{\mathbf{R}}_{abc}$ equals $\widehat{\mathbf{R}}_{dq0}$ since $\mathbf{T}_{dq0}\mathbf{T}_{dq0}^{-1}=\mathbf{I}$.  A similar procedure can be applied to transform the uncertain inductance values from the $abc$ frame to the $dq0$ frame.

\textit{Measurement uncertainty:} The measurements, as shown in Fig. 5 [59], may include noise $n$ multiplied by a weighting factor $w^n$ and possible calibration errors represented by a $\mathit{gain}$:  
\begin{equation}
\label{eq:uncertain_gain}
\mathit{gain}=1+w_m\delta_m
\end{equation}
where $\lvert\delta_m\rvert\leq1$ is a real random scalar and $w_m$ represents the percentage error of the measurements. It should be noted that the presence of noise is not considered in the robustness analysis, as it does not destabilize the system but only affects its performance. Therefore, its impact should be studied in robust performance analysis, which is not the focus of this paper.  Nevertheless, for completeness, noise is included in the developed model.
\begin{figure}[!t]
  \centering
  \includegraphics[width=\linewidth]{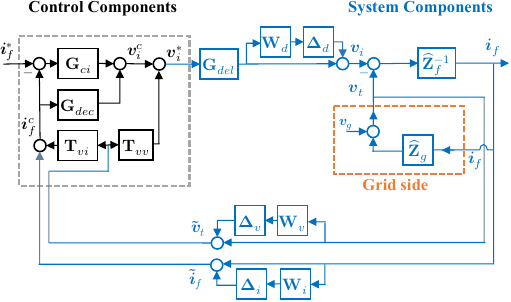}
  \caption{Model uncertainty augmented block diagram of the grid-tied VSC system.}
\end{figure}
The model in Fig. 5 represents measuring devices in $abc$ frame that should be transformed into the $dq0$ frame. In this paper, the following procedure is developed to address this issue. The measured voltage signals in the $abc$ frame are as follows:
\begin{equation}
\label{eq:abc_voltage_measurement_uncertainty}
\begin{bmatrix}
\widetilde{v}_a\\
\widetilde{v}_b\\
\widetilde{v}_c
\end{bmatrix}
=
\mathbf{W}_{abc}^{n}
\begin{bmatrix}
n_{v_a}\\
n_{v_b}\\
n_{v_c}
\end{bmatrix}
+
\mathbf{\Delta}_{abc}^{v}
\begin{bmatrix}
v_a\\
v_b\\
v_c
\end{bmatrix}
\end{equation}
where
\begin{equation}
\label{eq:abc_voltage_uncertainty_matrix}
\mathbf{\Delta}_{abc}^{v}
=
\begin{bmatrix}
1+w_{v_a}\delta_{v_a} & 0 & 0\\
0 & 1+w_{v_b}\delta_{v_b} & 0\\
0 & 0 & 1+w_{v_c}\delta_{v_c}
\end{bmatrix}
\end{equation}
\begin{equation}
\label{eq:abc_voltage_noise_weighting_matrix}
\mathbf{W}_{abc}^{n}
=
\begin{bmatrix}
w_{v_a}^{n} & 0 & 0\\
0 & w_{v_b}^{n} & 0\\
0 & 0 & w_{v_c}^{n}
\end{bmatrix}.
\end{equation}
By transforming (10) into $dq0$ frame, the following in (13) is obtained.
\begin{equation}
\label{eq:dq0_voltage_measurement_uncertainty}
\begin{aligned}
\begin{bmatrix}
\widetilde{v}_d\\
\widetilde{v}_q\\
\widetilde{v}_0
\end{bmatrix}
&=
\mathbf{T}_{dq0}\mathbf{W}_{abc}^{n}
\begin{bmatrix}
n_{v_a}\\
n_{v_b}\\
n_{v_c}
\end{bmatrix}
+
\mathbf{T}_{dq0}\mathbf{\Delta}_{abc_v}^{v}
\mathbf{T}_{dq0}^{-1}
\begin{bmatrix}
v_d\\
v_q\\
v_0
\end{bmatrix}
\\
&=
\mathbf{W}_{dq0}^{n}
\begin{bmatrix}
n_{v_a}\\
n_{v_b}\\
n_{v_c}
\end{bmatrix}
+
\mathbf{\Delta}_{dq0}^{v}
\begin{bmatrix}
v_d\\
v_q\\
v_0
\end{bmatrix}.
\end{aligned}
\end{equation}
Unlike $\mathbf{\Delta}_{abc}^{v}$, which is a diagonal matrix as presented in (11), $\mathbf{\Delta}_{dq0}^{v}$ is generally not diagonal according to (13). However, for simplicity, if the measuring devices are assumed to be identical across three phases, meaning in (11), $w_{v_a}\delta_{v_a}=w_{v_b}\delta_{v_b}=w_{v_c}\delta_{v_c}$, then $\mathbf{\Delta}_{abc}^{v}$ equals $\mathbf{\Delta}_{dq0}^{v}$ since $\mathbf{T}_{dq0}\mathbf{T}_{dq0}^{-1}=\mathbf{I}$. Also, for such identical measuring devices, the weighting factors across phases are such that $w_{v_a}=w_{v_b}=w_{v_c}$. The noise distributions $n_v$ are also assumed to be the same across three phases. In the $abc$ frame, a signal with additive noise of variance $\sigma^2$, when multiplied by a weighting factor $w$, results in noise with variance $w^2 \sigma^2$. When this signal is transformed to the $dq0$ frame using the orthonormal transformation $\mathbf{T}_{dq0}$, the noise retains its structure and is still multiplied by the same factor $w$, yielding a variance of $w^2 \sigma^2$ in the $dq0$ frame. A similar procedure can be followed for the current. 

\textit{Unmodeled dynamics:} This type of uncertainty can be due to factors such as modeling simplifications and/or unknown system dynamics. Unmodeled dynamics often are negligible at low frequencies but tend to amplify at higher frequencies, typically represented by a high-pass filter as follows [11]:
\begin{equation}
\label{eq:dynamic_uncertainty_weight}
\mathbf{W}_d(s)
=
\begin{bmatrix}
\dfrac{w_d s}{s+\omega_c} & 0\\[6pt]
0 & \dfrac{w_d s}{s+\omega_c}
\end{bmatrix}
\end{equation}
where $\omega_c$ denotes the cutoff frequency beyond which this uncertainty becomes significant, and $w_d$ represents its magnitude. The uncertainty associated with the unmodeled dynamics in each channel is modeled via $\lvert\delta_d\rvert\leq1$. 

Fig. 6 shows the developed model-uncertainty-augmented representation of the IBR-connected system using transfer function block diagrams where the transfer function blocks are $2\times2$ matrices. $\mathbf{G}_{ci}$ represents the PI current controller, and $\mathbf{G}_{dec}$ denotes the $dq0$ axis decoupling terms in the controller, which are constructed by following the procedure explained in [60]. $\mathbf{G}_{del}$ represents the aggregated delay of the controller digitalization plus PWM switching. Note that the delay block is modeled according to [61].  $\mathbf{Z}_{f}$ represents the transfer function of the inverter filter.   

According to [62], the interconnection of a stable VSC to a stable grid is stable if a feedback system with the loop transfer function of $\mathbf{G}(s) =\mathbf{Z}_{g}(s)\mathbf{Y}_{\mathrm{VSC}}$ is stable. In Fig. 6, the relationship between the output (or terminal) current $i_f$ and the terminal voltage $\boldsymbol{v}_t$ and $\boldsymbol{i}_f^*$ can be expressed as: 
\begin{equation}
\label{eq:vsc_current_relation}
\boldsymbol{i}_f(s)
=
\mathbf{G}_{ii^*}\boldsymbol{i}_f^*
-
\mathbf{Y}_{\mathrm{VSC}}(s)\boldsymbol{v}_t(s)
\end{equation}
where $\mathbf{G}_{ii^*}$ is the transfer function between the terminal current and its reference, and $\mathbf{Y}_\mathrm{VSC}$ is the equivalent admittance of the VSC. Also, $\mathbf{Z}_g$ is as follows:
\begin{equation}
\label{eq:grid_impedance_matrix}
\mathbf{Z}_g(s)
=
\begin{bmatrix}
L_gs+R_g & -\omega_s L_g\\
\omega_s L_g & L_gs+R_g
\end{bmatrix}.
\end{equation}
\begin{figure}[!t]
  \centering
  \includegraphics[width=\linewidth]{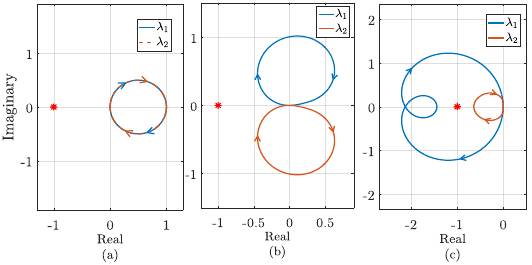}
  \caption{Nyquist plot of (a) nominal system for Case I, (b) perturbed system for $k_{11}=0.98$, (c) perturbed system for $k_{11}=0.92$.}
\end{figure}
\subsection{Case studies}
This section presents numerical results showing that robustness assessment using the GNC framework is not reliable. Then, it illustrates how proper robustness assessment can be performed using the $\mu$-analysis. 

Assume that the nominal loop transfer function of the system is $\mathbf{G}^{0}$. The characteristic loci (i.e., the Nyquist plot) of $\mathbf{G}^{0}$, denoted by $\mathbf{G}^{0}_\mathrm{N}$, crosses the negative real axis at frequency $\omega_c^0$ with a magnitude of $\lvert\mathbf{G}^{0}_\mathrm{N}(j\omega_c^0)\rvert$. Let $\mathbf{G}^{'}$ represent the uncertain system containing model uncertainties. As required by the GNC framework, the uncertain system is represented in terms of a multiplicative uncertainty block $\mathbf{U}$ and the nominal system as follows:
\begin{equation}
\label{eq:transformed_frequency_response}
\mathbf{G}'(j\omega)
=
\mathbf{U}(j\omega)\mathbf{G}^{0}(j\omega)
\end{equation}
where
\begin{equation}
\label{eq:transformation_matrix_from_frequency_responses}
\mathbf{U}(j\omega)
=
\mathbf{G}'(j\omega)
\left[\mathbf{G}^{0}(j\omega)\right]^{-1}.
\end{equation}
As discussed earlier, in the GNC framework, the GM analysis assumes that $\mathbf{U}$ is a scalar matrix of the form $\mathbf{U}=\begin{bmatrix}u & 0\\0 & u\end{bmatrix}=u\mathbf{I}_{2\times2}$. Consequently, the perturbed system due to uncertainty is considered as follows:  
\begin{equation}
\label{eq:uniformly_perturbed_frequency_response}
\mathbf{G}^{\mathrm{P,Uniform}}
=
\mathbf{K}\mathbf{G}^{0}
=
\begin{bmatrix}
k & 0\\
0 & k
\end{bmatrix}
\mathbf{G}^{0}
=
\begin{bmatrix}
k g_{11}^{0} & k g_{12}^{0}\\
k g_{21}^{0} & k g_{22}^{0}
\end{bmatrix}
\end{equation}
where $\mathbf{K}=\operatorname{mag}(\mathbf{U})$. This section will show that the assumption of $\mathbf{U}$ being a scalar matrix is valid only for very special cases of uncertainty in an IBR system. For general cases, the uncertainty block takes the form $\mathbf{U}=\begin{bmatrix}u_{11} & u_{12}\\u_{21} & u_{22}\end{bmatrix}$. This difference has several important consequences for robustness analysis, as follows:
\begin{itemize}
    \item The GNC framework assumes $\mathbf{U}$ is a scalar matrix (i.e., $\displaystyle \mathbf{U}=\begin{bmatrix}u & 0\\0 & u\end{bmatrix}=u\mathbf{I}_{2\times2}$), and the uncertainty is characterized by a single scalar $u$. The traditional gain and phase margin concepts in the GNC framework treat the effects of magnitude and phase variations in u separately. When $\mathbf{U}$ is not a scalar matrix, it involves four scalars $u_{ij}$.The resulting directionality and coupling effects further limit the suitability of analyzing the impact of $\mathbf{U}$ on robustness using independent evaluations of its gain and phase variations.
    \item To implement the GM and PM concept from the GNC framework, the perturbed system corresponding to the general uncertainty case becomes as follows:
    \begin{equation}
    \label{eq:perturbed_frequency_response}
    \mathbf{G}^{\mathrm{P}}
    =
    \mathbf{K}\mathbf{G}^{0}
    =
    \begin{bmatrix}
    k_{11} & k_{12}\\
    k_{21} & k_{22}
    \end{bmatrix}
    \begin{bmatrix}
    g_{11}^{0} & g_{12}^{0}\\
    g_{21}^{0} & g_{22}^{0}
    \end{bmatrix}
    \end{equation}
    where $k_{ij}=\operatorname{mag}(u_{ij})$. In (19), the Nyquist plot of
    $\mathbf{G}^{\mathrm{P,Uniform}}$, denoted by
    $\mathbf{G}_{\mathrm{N}}^{\mathrm{P,Uniform}}$, is a linear function of
    $\mathbf{K}$ because $\mathbf{K}$ is a scalar matrix, but in (20), the
    Nyquist plot of $\mathbf{G}^{\mathrm{P}}$, denoted
    $\mathbf{G}_{\mathrm{N}}^{\mathrm{P}}$, is a nonlinear function of
    $\mathbf{K}$. In this section, it will be demonstrated that for a uniform
    uncertainty scenario in an IBR system, the gain margin in (19) can correctly
    predict the behavior of the perturbed system
    $\mathbf{G}^{\mathrm{P,Uniform}}$. This means if $\omega_c^0$ is the phase
    crossover frequency of $\mathbf{G}_{\mathrm{N}}^0$, scaling by $k$ causes
    the Nyquist plot $\mathbf{G}_{\mathrm{N}}^{\mathrm{P,Uniform}}$ to cross the
    critical point $-1+j0$ at the same frequency $\omega_c^0$ and
    $\left|\mathbf{G}_{\mathrm{N}}^{\mathrm{P,Uniform}}(j\omega_c^0)\right|=1$.
    In contrast, for the general uncertainty case described by (20), the
    Nyquist plot $\mathbf{G}_{\mathrm{N}}^{\mathrm{P}}$ may not have predictable
    behavior. This occurs because the $\mathbf{K}$ matrix of (20) causes a
    nonlinear change from $\mathbf{G}_{\mathrm{N}}^0$ to
    $\mathbf{G}_{\mathrm{N}}^{\mathrm{P}}$. Therefore, gain margin may not be
    able to predict different aspects of the behavior of
    $\mathbf{G}_{\mathrm{N}}^{\mathrm{P}}$, such as the value of
    $\mathbf{G}_{\mathrm{N}}^{\mathrm{P}}(j\omega_c^0)$.
\end{itemize}

\subsubsection*{Case I: Numerically constructed system:} This case is not related to the IBR system of Fig. 3. Rather, it is a numerically constructed system to show an extreme scenario where a nominal system with infinite gain margin, determined based on GNC, becomes unstable for a small non-uniform uncertainty. The nominal system is as follows:
\begin{equation}
\label{eq:nominal_transfer_function_matrix}
\mathbf{G}^{0}(s)
=
\begin{bmatrix}
\dfrac{101s+203}{s^{2}+5s+6}
&
\dfrac{100}{s+3}
\\[8pt]
-\dfrac{100}{s+3}
&
\dfrac{-99s-197}{s^{2}+5s+6}
\end{bmatrix}.
\end{equation}
Fig. 7-(a) shows the GNC of this transfer matrix, where the gain margin is infinite. Consider an output uncertainty of
$\mathbf{K}=\begin{bmatrix}k_{11} & 0\\0 & 1\end{bmatrix}$.
Fig. 7-(b) and Fig. 7-(c) show the impact of this uncertainty for $k_{11}=0.98$ and $k_{11}=0.92$, respectively. According to Fig. 7-(c), the system becomes unstable for such a small non-uniform uncertainty, which contrasts with the predicted infinite GM predicted by the GNC framework. Also, Fig. 7-(a) to Fig. 7-(c) show that due to the nonlinear behavior of the Nyquist plot of the uncertain system with non-uniform uncertainty $\mathbf{K}\mathbf{G}^0$, a small change in the value of $k_{11}$ causes a significant change in the Nyquist plots in an unpredictable way. The remaining cases are related to Fig. 3.

\subsubsection*{\textit{Case II: Uniform uncertainty in} $\mathbf{Z}_g$}
As shown in Fig. 3, a $10$ MVA IBR is connected to the grid where the strength of the system defined by short circuit ratio (SCR) at the terminal of IBR is $\mathrm{SCR=2}$, with an impedance angle of $50^\circ$. The IBR is injecting active and reactive powers of $P=0.8$ p.u and $Q=0.2$ p.u, respectively. In this case the uncertainties in $\mathbf{Y}_\mathrm{VSC}$ of the inverter are ignored. To create a uniform uncertainty scenario in $\mathbf{Z}_g$, it is assumed the perturbation is in such a way that the impedance angle remains unchanged. In other words, in (16), the ratio of $L_g/R_g$ does not change due to the perturbation. To study the behavior of $\mathbf{G}_\mathrm{N}^\mathrm{P}$ as it crosses the critical point of $-1+j0$, the parameters of $\mathbf{Z}_g$ are changed until $\mathbf{G}_\mathrm{N}^\mathrm{P}$ crosses $-1+j0$. A possible perturbation scenario involves a change in the power grid (e.g., topology changes and outages) that causes both equivalent inductance $L_g$ and resistance $R_g$ to increase by $54\%$, leading to a reduction of the SCR to $1.3$. Fig. 8-(a) shows $\mathbf{G}_N^0$. Based on $\mathbf{G}_N^0$, the system has a $\mathrm{GM}=1.53$, and $\mathrm{PM}=19.83^\circ$, where the phase crossover frequency is $\omega_c^0=41.37$ rad/s. Fig. 8-(b) shows the $\mathbf{G}_\mathrm{N}^\mathrm{P}$ plot under the abovementioned uniform uncertainty, and $k_{ij}$ values are as follows:
$$
\mathbf{K}(j\omega_c^{0})
=
\begin{bmatrix}
1.53 & 0\\
0 & 1.53
\end{bmatrix}
=
1.53\mathbf{I}_{2\times2}.
$$
According to Fig. 8-(a) and (b),  $\mathbf{G}_\mathrm{N}^\mathrm{0}$ is linearly expanded to $\mathbf{G}_\mathrm{N}^\mathrm{P}$. For example, $\mathbf{G}_\mathrm{N}^\mathrm{P}$ crosses $-1+j0$ at the same frequency that $\mathbf{G}_\mathrm{N}^\mathrm{0}$ crosses the negative real axis (i.e., $\omega_c^0=41.37$ rad/s). This means, as expected, under this uniform uncertainty the GNC predicts/estimates the value of GM correctly.
\begin{figure}[!t]
  \centering
  \includegraphics[width=\linewidth]{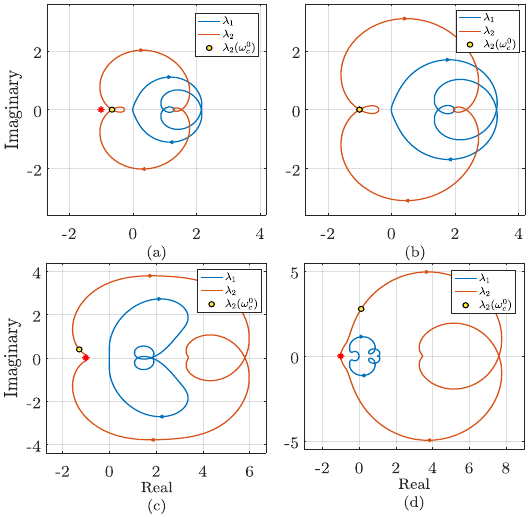}
  \caption{(a) Nyquist plot of nominal system for Case II, (b), Nyquist plots of the perturbed system for Case II, (c) Nyquist plot of the perturbed system for Case III, (d), Nyquist plot of the perturbed system for Case IV.}
\end{figure}
\begin{figure}[!t]
  \centering

  \includegraphics[width=\linewidth]{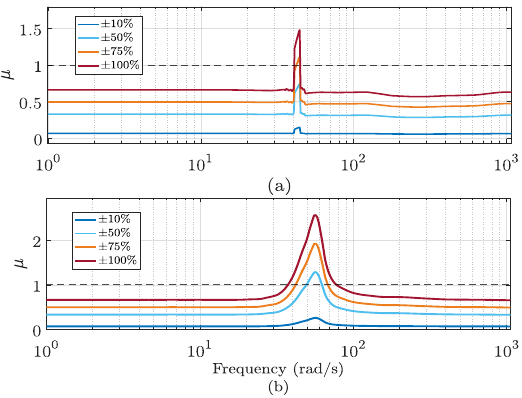}
  \caption{$\mu$‑analysis results for perturbations in $\mathbf{Z}_g$ that cause (a) $L_g/R_g$ ratio remains constant like Case II (b) $L_g/R_g$ changes like Case III.}

  \vspace{0.5em}

  \includegraphics[width=\linewidth]{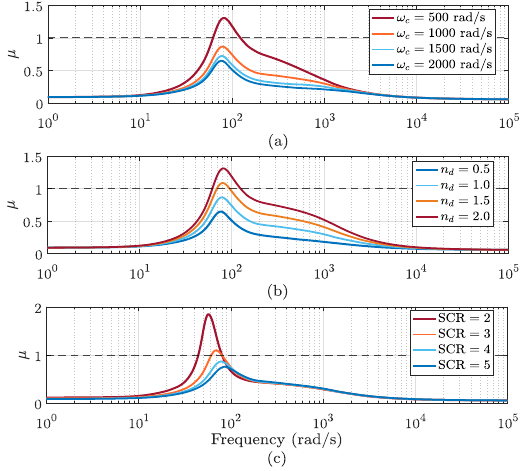}
  \caption{$\mu$‑analysis results for perturbations in both $\mathbf{Z}_g$ and $\mathbf{Y}_\mathrm{VSC}$ with different levels of unmodeled dynamic uncertainty.}

\end{figure}
\subsubsection*{\textit{Case III: non-uniform uncertainty in} $\mathbf{Z}_g$}
In this case, the nominal system is the same as Case II. However, it is assumed that due to the topology changes in the grid, $L_g/R_g$  ratio does not remain the same. To study the behavior of $\mathbf{G}_\mathrm{N}^\mathrm{P}$ as it crosses the critical point of $-1+j0$, parameters of $\mathbf{Z}_g$ are changed until $\mathbf{G}_\mathrm{N}^\mathrm{P}$ crosses $-1+j0$. A possible perturbation scenario involves a change in the power grid (e.g., topology changes and outages) that causes the equivalent grid inductance $L_g$ increases by $148\%$, while $R_g$ remains unchanged, which leads to $\mathrm{SCR}=1$.  In this case, the perturbation is as follows:
$$
\mathbf{K}(j\omega_c^{0})
=
\begin{bmatrix}
1.7704 & 2.3003\\
0.1587 & 2.0195
\end{bmatrix}.
$$
This shows the predicted GM by GNC is not reliable, as it predicted a GM of $1.53$, meaning
$\mathbf{K}(j\omega_c^0)=1.53\mathbf{I}_{2\times2}$, but
$\mathbf{K}(j\omega_c^0)$ is as above, which cannot be presented by a scalar value as GM.
Also, according to Fig. 8-(a) and (c), $\mathbf{G}_{\mathrm{N}}^0$ is nonlinearly expanded to
$\mathbf{G}_{\mathrm{N}}^{\mathrm{P}}$. For instance, GM calculation based on GNC expects the
value corresponding to $\omega=\omega_c^0=41.37\ \mathrm{rad/s}$ on
$\mathbf{G}_{\mathrm{N}}^{\mathrm{P}}$ to be $-1+j0$ as it assumes a linear expansion of
$\mathbf{G}_{\mathrm{N}}^0$, but as shown in Fig. 8-(c) by a yellow circle, it is not $-1+j0$.

\textit{Case IV: Uncertainty in both $\mathbf{Y}_{\mathrm{VSC}}$ and $\mathbf{Z}_g$}
This case considers uncertainties in $\mathbf{Z}_g$ and uncertainties in
$\mathbf{Y}_{\mathrm{VSC}}$. To observe the behavior of
$\mathbf{G}_{\mathrm{N}}^{\mathrm{P}}$ as it crosses the critical point of
$1+j0$, parameters of $\mathbf{Z}_g$ are changed until
$\mathbf{G}_{\mathrm{N}}^{\mathrm{P}}$ crosses $-1+j0$. A possible
perturbation scenario involves changes in the power grid (e.g., topology
changes and outages) that reduce the SCR to 1.05, with $R_g$ increased by
$186\%$ and $L_g$ decreased by $38\%$. In addition, an $18\%$ uncertainty
is introduced to $R_f$, $L_f$, and $T_d$. In this case, $\mathbf{K}$ is as
follows:
\[
\mathbf{K}(j\omega_c^0)
=
\begin{bmatrix}
2.0722 & 3.4136\\
0.2430 & 1.4288
\end{bmatrix}.
\]
This shows the predicted GM by GNC is not reliable as it predicted a GM of
1.53, meaning $\mathbf{K}(j\omega_c^0)=1.53\mathbf{I}_{2\times2}$, but
$\mathbf{K}(j\omega_c^0)$ is as above, which cannot be presented by a scalar
value as GM. According to Fig. 8-(a) and Fig. 8-(d),
$\mathbf{G}_{\mathrm{N}}^0$ is nonlinearly expanded to
$\mathbf{G}_{\mathrm{N}}^{\mathrm{P}}$. For instance, GM calculation based
on GNC expected the value of $\mathbf{G}_{\mathrm{N}}^{\mathrm{P}}$ at
$\omega=\omega_c^0=41.37\ \mathrm{rad/s}$ to be $-1+j0$, as it assumes a
linear expansion of $\mathbf{G}_{\mathrm{N}}^0$. However, as shown in Fig. 8-(d) by the yellow circle, the value of $\mathbf{G}_\mathrm{N}^\mathrm{P}$ corresponding to $\omega = \omega_c^0=41.37$ rad/s is not $-1+j0$. 

\begin{figure}[!t]
  \centering
  \includegraphics[width=\linewidth]{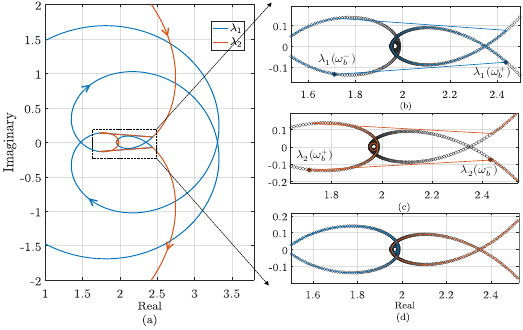}
  \caption{Labeling issue of the Nyquist plot of $\mathbf{G}(s)$.}
\end{figure}

\subsubsection*{\textit{Case IV: MIMO robustness assessment using $\mu$‑analysis}}
Previous cases demonstrated the limitations of the GNC framework for robustness assessment. As discussed before, the proper way of robustness assessment for MIMO systems is to use MIMO robustness analysis tools such as $\mu$‑analysis. In this 
section, several uncertainty scenarios are considered to demonstrate how proper robustness analysis can be performed using $\mu$-analysis for the IBR system of Fig. 3 using the uncertainty models of $\mathbf{Y}_\mathrm{VSC}$ developed in Fig. 6 and uncertainties of $\mathbf{Z}_g$.

Fig. 9 shows the $\mu$ analysis results for perturbations in $\mathbf{Z}_g$, while uncertainties in $\mathbf{Y}_\mathrm{VSC}$ are neglected.  Fig. 9-(a) considers the perturbation scenario corresponding to Case II, where the ratio $L_g/R_g$ remains constant. Fig. 9-(b) shows uncertainty scenarios like Case III, where $L_g/R_g$ does not remain constant. As stated before, in $\mu$ analysis, the distance between the $\mu$ value and the line  $\mu=1$ represents the robustness margin. The farther the $\mu$ value lies below 1, the more robust the system is. 
For example, in Fig. 9-(a), the plot represents the system with $10\%$ uncertainty in $R_g$ and $L_g$ is more robust than the plot representing $50\%$ uncertainty in $R_g$ and $L_g$. It should be noted that the $\mu$ analysis framework readily incorporates possible available information about system uncertainties into the analysis. For instance, in Fig. 11-(a), the $\mu$ analysis is
performed under the condition that $L_g/R_g$ remains constant, whereas this restriction is removed in Fig. 9-(b). Such flexibility cannot be directly accommodated into $\mathbf{G}(s)=\mathbf{Z}_g(s) \mathbf{Y}_\mathrm{VSC}(s)$ within the conventional GNC framework. 

Fig. 10 shows the structured singular value $\mu$ as a function of frequency, considering all sources of uncertainty in both $\mathbf{Y}_\mathrm{VSC}$ and $\mathbf{Z}_g$. These include measurement uncertainty ($5\%$ 
for both $\boldsymbol{v}_t$ and $\boldsymbol{i}$), parametric uncertainty ($5\%$ for all parameters), and unmodeled dynamic uncertainties with varying weights at locations shown in Fig. 6. 

The $\mu$ analysis, as presented in Fig. 9 and Fig. 10, provides valuable insights for designing targeted mitigation strategies, such as refining the controller or improving model fidelity, particularly near the frequencies where the structured singular value $\mu$ approaches $1$.

\subsubsection*{\textit{Case V: Labeling issue of the Nyquist plots}}
This section presents a scenario where mislabeling of eigenvalues leads to discontinuities in the eigenvalue loci. For the $2\times2$ transfer function matrix $\mathbf{G}(s)$, the eigenvalues at each frequency $\omega$ are obtained as the roots of $\det\!\left(\lambda\mathbf{I}-\mathbf{G}(j\omega)\right)$. In Fig. 11-(b) to Fig. 11-(d), the calculated eigenvalues at each frequency
are shown by circles. One possible approach, which MATLAB also adopts, is to label these eigenvalues as
\[
\lambda_1
=
\frac{\operatorname{tr}\!\left(\mathbf{G}(j\omega)\right)}{2}
+
\frac{1}{2}
\sqrt{
\operatorname{tr}^{2}\!\left(\mathbf{G}(j\omega)\right)
-
4\det\!\left(\mathbf{G}(j\omega)\right)
}
\]
\[
\lambda_2
=
\frac{\operatorname{tr}\!\left(\mathbf{G}(j\omega)\right)}{2}
-
\frac{1}{2}
\sqrt{
\operatorname{tr}^{2}\!\left(\mathbf{G}(j\omega)\right)
-
4\det\!\left(\mathbf{G}(j\omega)\right)
}.
\]

The term
$d(j\omega)=\operatorname{tr}^{2}\!\left(\mathbf{G}(j\omega)\right)
-4\det\!\left(\mathbf{G}(j\omega)\right)$ is, in general, a complex number,
expressed as $d=re^{j\theta}$. Its square root can be expressed as
$\sqrt{d}=\sqrt{re^{j\theta}}=\sqrt{r}\cdot e^{j(\theta/2)}$.
In this calculation, the principal branch of the complex square root is used,
which restricts the argument $\theta$ of any complex number to the range
$(-\pi,+\pi]$. When $d(j\omega)$ crosses the negative real axis as the
frequency increases, the phase angle $\theta$ changes from a value slightly
less than $180^\circ$ (say $178^\circ$) to a value slightly greater than
$180^\circ$ (say $182^\circ$), which, under the principal branch convention,
is equivalent to $-178^\circ$. Consequently, $\theta/2$ jumps from
$+89^\circ$ to $-89^\circ$, producing a sudden change in the values of
$\lambda_1$ and $\lambda_2$ as defined above. This results in a discontinuity
in the eigenvalue trajectories.

In the Nyquist plot of the IBR system, the above incorrect labeling approach
is also observed in Fig. 11-(b) and Fig. 11-(c), which show the labeling of
$\lambda_1$ and $\lambda_1$ loci. Around the breaking frequency
$\omega_b=17.92\ \mathrm{rad/s}$, it is observed that
$\sqrt{d(j\omega_b^-)}=0.72\angle175.38^\circ$, but
$\sqrt{d(j\omega_b^+)}=0.72\angle-4.63^\circ$, where $\omega_b^-$ and
$\omega_b^+$ represent frequencies slightly smaller and slightly bigger than
the breaking frequency, respectively. This causes $\lambda_2$ jumps from
$\lambda_2(\omega_b^-)=2.4402+j0.0765$ to
$\lambda(\omega_b^+)=1.7235+j0.1346$. This abrupt change leads to a sharp edge
in the eigenvalue locus, clearly indicating a discontinuity. In Fig. 11-(a),
this issue is addressed by labeling the eigenvalues according to the
minimum-distance proximity criterion. As shown in Fig. 11-(d), the continuity of the labeled eigenvalues is maintained.

\section{Conclusions}
This paper presented both theoretical and numerical analyses to demonstrate the limitations and misconceptions of the GNC in evaluating the stability of grid‑connected IBRs. It showed that the GNC‑based robustness analysis is valid only under uniform uncertainties. In practice, since real‑world uncertainties are typically non‑uniform, the GNC is not a reliable measure of robustness. Therefore, unlike SISO systems, where gain and phase margins provide simple and effective means to assess a system’s vulnerability to instability, this relationship generally does not hold for MIMO systems. Because of non‑uniform uncertainty, the gain and phase margin values obtained from the GNC framework are not reliable. This means that, unlike SISO system that the distance of the Nyquist plots from the critical point $-1+j0$ is an indicator of robustness, in MIMO systems this does not hold, as non-uniform disturbances affect the Nyquist plots nonlinearly. To demonstrate proper robust stability assessment for MIMO IBR systems, the paper developed a model uncertainty-augmented representation and used it for robustness evaluation using $\mu$‑analysis. 

\section{Acknowledgment} 
The authors would like to thank Dr. Ali Saberi from Washington State University and Dr. Anton A. Stoorvogel from the University of Twente for their valuable discussions and insights during the preparation of the paper.

%
%
\vspace{-12cm} 

\begin{IEEEbiography}[
  {\includegraphics[width=1in,height=1.25in,clip,keepaspectratio]{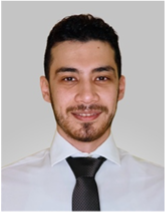}}
]{Hassan Yazdani}
(S’25) is currently pursuing a Ph.D. in Electrical Engineering at Washington State University, Pullman, WA, USA. His research interests include the dynamics, control, and stability analysis of inverter-based power systems, as well as optimization and machine learning applications in power grids.
\end{IEEEbiography}

\vspace{-13cm} 

\begin{IEEEbiography}[
  {\includegraphics[width=1in,height=1.25in,clip,keepaspectratio]{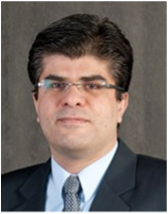}}
]{Saeed Lotfifard}
(S’08–M’11–SM’17) received his Ph.D. degree in electrical engineering from Texas A\&M University, College Station, TX, in 2011. Currently, he is an associate professor at Washington State University, Pullman. His research interests include stability, protection, and control of inverter-based power grids. He serves as an associate editor for the \textsc{IEEE Transactions on Power Delivery}.
\end{IEEEbiography}

\end{document}